\documentclass[copyright,creativecommons]{eptcs}

\usepackage{subfigure}
\usepackage{listings}
\usepackage{times}
\usepackage{epsfig}

\lstdefinelanguage{OOPL}{  sensitive=false,  keywords={Roles, Facts, Effects, Sanctions, Brute, Effect, Sanction, Rules, CountsAs, return, does},  otherkeywords={Norm-level}}

\newcounter{lst}[section]

\lstnewenvironment{listing}[3]{
  \noindent        
  \refstepcounter{lst}         
  \label{code:#1}  
\begin{tabular}{p{\textwidth}} \hline  {\small \textbf{Code  fragment \arabic{section}.\arabic{lst}} #2} \\  \end{tabular} 
\lstset{language=#3,          
  basicstyle=\footnotesize\ttfamily,    
  xleftmargin=10pt,    
  mathescape=true,    
  frame=tb,    
  numbers=right,    
  numberstyle=\tiny,     
  stepnumber=1,     
  numbersep=-5pt}}{}

\lstdefinelanguage{Gwendolen}{%
    morekeywords={Plans,Initial,Beliefs,Goals,name},
 literate= {<-}{{$\leftarrow$}{$\:$}}2
           {.B}{{${\cal B}$}}2
           {.G}{{${\cal G}$}}2
           {True}{{$\top$}}2
           {(perform)}{}0
}

\newcommand{\SetGWENDOLEN}{%
  \lstset{%
    language=Gwendolen,
    basicstyle=\small\ttfamily,
    commentstyle=\normalfont\upshape,
    morekeywords={Plans,Initial,Beliefs,Goals,name},
    xleftmargin=0.2cm,
    aboveskip=0.3cm,
    belowskip=0cm,
    showstringspaces=false,
    flexiblecolumns=true,
    mathescape=true,
 }
}

\lstnewenvironment{gwendolenlst}[1][]{\lstset{frame=#1}\SetGWENDOLEN}{}

\title{Agent Based Approaches to Engineering Autonomous Space Software\thanks{Work funded by EPSRC grants EP/F037201/1 and EP/F037570/1}}
\author{Louise A. Dennis\institute{Department of Computer Science, University of Liverpool,
  UK }\email{L.A.Dennis@liverpool.ac.uk} \and  Michael Fisher\institute{Department of Computer Science, University of Liverpool,
  UK } \and Nicholas Lincoln\institute{Department of Engineering, University of
  Southampton, UK} \and Alexei Lisitsa\institute{Department of Computer Science, University of Liverpool,
  UK } \and Sandor M. Veres\institute{Department of Engineering, University of
  Southampton, UK}}

\usepackage{eass}

\begin{document}
\maketitle

\begin{abstract}
  Current approaches to the engineering of space software such as
  satellite control systems are based around the development of
  feedback controllers using packages such as MatLab's Simulink
  toolbox.  These provide powerful tools for engineering real time
  systems that adapt to changes in the environment but are limited
  when the controller itself needs to be adapted.

  We are investigating ways in which ideas from temporal logics and
  agent programming can be integrated with the use of such control
  systems to provide a more powerful layer of autonomous decision
  making.  This paper will discuss our initial approaches to the
  engineering of such systems.
\end{abstract}

\section{Introduction}

Modern control systems are limited in their ability to react flexibly
and autonomously to changing situations. The limiting factor is the
complexity inherent in analysing situations where many variables are
present. There are many complex, real-world, control systems but we
are primarily interested in the (autonomous) control of satellite
systems.

Consider the problem of a single satellite attempting to maintain a
geostationary orbit.  Current satellite control systems maintain
orbits using feedback controllers.  These implicitly assume that any
errors in the orbit will be minor and easily corrected.  In situations
where more significant errors occur, for example caused by a thruster
malfunction, it is desirable to modify or change the controller.  The
complexity of the decision task is a challenge to standard approaches,
and has led, for example, to complex, evolutionary control
systems. These become very difficult to
understand. 

We approach the problem from the perspective of rational
agents~\cite{WooldridgeRao99:book}.  We consider a satellite to be an
{\em agent} which consists of a discrete (rational decision making)
part and a continuous (calculation) part.  The discrete part uses the
{\em Belief-Desire-Intention} (BDI) theory of agency~\cite{rao:95b}
and governs high level decisions about when to generate new feedback
controllers.  The continuous, calculational part is used to derive
controllers and to calculate information from continuous data which
can be used in the decision making process; this part can be viewed as
a hybrid system.

\section{Architecture}

\begin{figure}[htb]
\begin{center}
\includegraphics[width=3.95in]{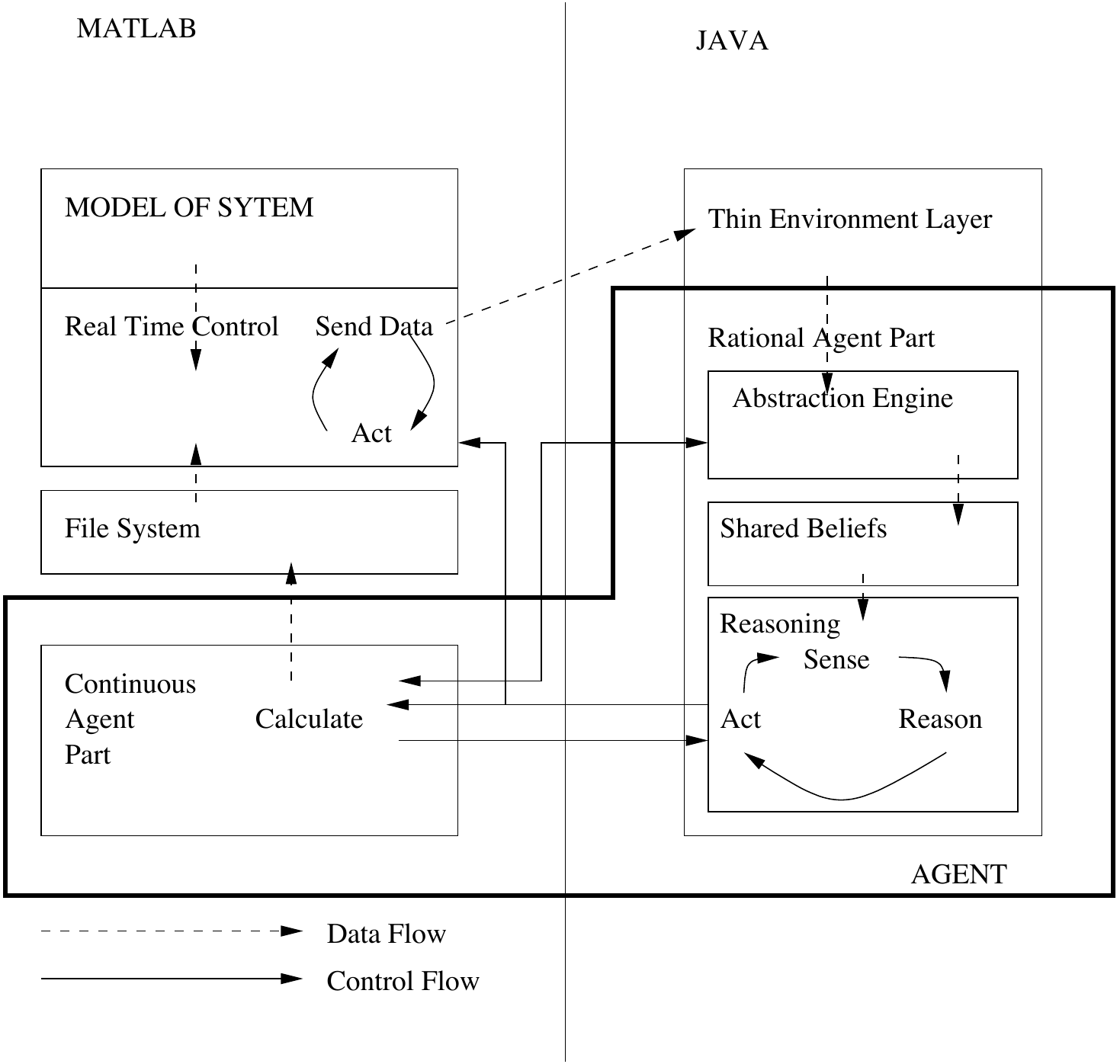}
\caption{Implemented Hybrid Agent Architecture}
\label{fig:arch}
\end{center}
\end{figure}

Our prototype system is shown in Fig.~\ref{fig:arch}.  We have
implemented a simulated environment and real time satellite control
system in MatLab using the Simulink tool kit.  The {\em continuous
  agent part} is also implemented in MatLab.  MatLab has no easy provision
for threaded execution which forces us to use separate instances for
the Real Time aspects (i.e. the feedback controller and simulated
environment) and for the Continuous Agent part.  The agent also
contains a {\em discrete agent part} which is currently implemented in
the \gwendolen{} agent programming language\footnote{The choice of
  language was dictated entirely by convenience.  It is a subject for
  further work to examine more widely used BDI-languages and evaluate
  which is most appropriate for the system.}.
\gwendolen{}~\cite{dennis08:_gwend} is implemented on top of \java.

The real time control system sends information (which may be
pre-processed) to the agent part of the system.  When it acts, the
discrete part of the agent may either cause the continuous agent part
to perform some calculation (and wait for the results) or it may send
an instruction to the real time control system to alter its
controller.  Since the new controller has been created ``on the fly''
by the continuous part, some aspects of this controller are stored in
the shared file system (accessed by both MatLab processes).

The discrete agent part is divided into an {\em abstraction engine}
which takes continuous data supplied by the satellite simulation and
transforms this data into discrete shared beliefs which are accessed
by a {\em reasoning engine} which makes decisions about how to
behave.  The discrete part is split in two because reasoning
is comparatively slow compared to the flow of data coming in from the
simulation.  It can become ``clogged'' up with the need to react to
changing information if it tries to perform both the abstraction tasks
and the reasoning tasks at once.  The separation of abstraction and reasoning is both
theoretically clean and practical at an implementational level.

\section{BDI Programming Aspects}

The architecture lets us represent the high-level decision making
aspects of the program in terms of the beliefs and goals of the agent
and the events it observes.  So, for instance, when the agent observes
the event that the satellite is in a new position (information relayed
to it by the real time controller) it can call on the continuous part
to calculate whether this position is within acceptable bounds of the
desired orbit (i.e. whether the existing real-time controller is
capable of maintaining the orbit).  If, as a result of this, it gains
a belief that the satellite has strayed from the orbit it can request
the continuous part to calculate a new path for the satellite to
follow using techniques described in~\cite{PaperRefNumber}.

Similarly, if the satellite has strayed from its bounds, the discrete
agent part can examine its beliefs about the current status of the
thrusters and, if necessary, instruct the continuous part to generate
a new feedback controller which takes into account any malfunctions or
inefficiencies in the thrusters.

Such programs can be expressed compactly in the BDI-programming style
without the need for programming large decision trees to consider all
possible combinations of thruster status and satellite positions.
This should then reduce the probability of error in the
decision-making parts of the program and opens the possibility that
existing techniques for model checking such
programs~\cite{bordini08:_aeutom_verif_multi_agent_progr} can be
adapted to verify this part.

\subsection{Geostationary Orbit Case Study}
\SetGWENDOLEN

The agent code for the geostationary orbit is shown in code fragments
\ref{code:geo_abs} and \ref{code:geo_reas}. Fragment~\ref{code:geo_abs} shows the
code for the abstraction engine.  Every time it ``perceives'' the
satellite position (\lstinline{stateinfo}) it calls upon MatLab to
calculate whether or not this position is within bounds
(\lstinline{comp_distance}) and then asserts and removes shared
beliefs appropriately.

The code is shown as a series of plans of the form
\lstinline*trigger:{guard} <- deeds* where the trigger is some event
observed by the agent, the guard is a set of facts that must be true
before the plan is activated and the deeds are a stack of deeds to be
executed.  \lstinline{+b} is the addition of a belief, \lstinline{b},
and \lstinline{-b} is the removal of the belief, \lstinline{b}.  In a
guard \lstinline{.B b} means that \lstinline{b} is believed.  
\medskip

\begin{listing}{geo_abs}{Geostationary Orbit Control (Abstraction Engine)}{Gwendolen}
+stateinfo(L1, L2, L3, L4, L5, L6) : 
   {.B proximity_to_centre(V1)} <- 
      comp_distance(L1, L2, L3, L4, L5, L6, Val),
      +proximity_to_centre(Val);

+proximity_to_centre(in) : {.B proximity_to_centre(out)} <- 
      -proximity_to_center(out),
      remove_shared(proximity_to_centre(out)),
      assert_shared(proximity_to_centre(in));

+proximity_to_centre(out) : 
   {.B proximity_to_centre(in), 
    .B stateinfo(L1, L2, L3, L4, L5, L6)} <-
      -proximity_to_centre(in),
      remove_shared(stateinfo(A1, A2, A3, A4, A5, A6)),
      assert_shared(stateinfo(L1, L2, L3, L4, L5, L6)), 
      remove_shared(proximity_to_centre(in)),
      assert_shared(proximity_to_centre(out));
\end{listing}
\vspace{2mm}

\noindent Fragment~\ref{code:geo_reas} reacts to the dynamic information
about whether the satellite is within bounds or not.  It may call a
MatLab function, \lstinline{plan_approach_to_centre} which returns the
name of a plan to move a satellite back within bounds.
\lstinline{apply_controls} and \lstinline{maintain_path} are actions
applied to the simulation of the satellite which apply a named plan,
or continue normal operation as appropriate.  The syntax
\lstinline{+!g} indicates the acquisition of a goal.
\medskip

\begin{listing}{geo_reas}{Geostationary Orbit Control}{Gwendolen}
+proximity_to_centre(out) : {True} <- 
     -proximity_to_centre(in), 
     +!get_to_centre;

+proximity_to_centre(in) : {True} <- 
     -proximity_to_centre(out), 
     maintain_path;

+!get_to_centre : 
   {.B proximity_to_centre(out), 
    .B stateinfo(L1, L2, L3, L4, L5, L6)} <-
     plan_approach_to_centre(P, locn(L1, L2, L3, L4, L5, L6)),
     +!try_execute(P) (perform);

+!try_execute(P) : {.B proximity_to_centre(out)} <- 
     apply_controls(P);
\end{listing}

\subsection{Decision and Control}
The important aspect of both the above example and the architecture in
general is that the (MatLab) control systems take care of the detailed
calculation of continuous functions (paths, etc), while the rational
agent takes care of high-level decisions about targets and plans. This
separation of concerns simplifies both parts and avoids the problems
associated with large, opaque, complex, adaptive and evolutionary
control systems.

\section{Future Work}

We are currently working on our prototype system and case study which will
allow us to make comparisons of this agent approach to autonomous
decision-making in satellite systems to approaches based on finite
state machines and standard control.  We also are interested in
investigating the use of temporal logic and model checking to generate
forward planning capabilities for the agent along the lines of those
investigated by Kloetzer and Belta~\cite{KloetzerBelta08:TAC}.  We aim
to explore the possibility of using model checking to verify aspects
of the agent's behaviour. Given that we already have a formal
verification system for \gwendolen{}
agents~\cite{bordini08:_aeutom_verif_multi_agent_progr}, there is a
strong possibility that we can extend this to cope with (abstractions
of) the continuous part. As the diagram below shows, we already have
model-checking tools for the discrete/finite parts. Our interest now
is how far such techniques can be extended to account for other
aspects of the agent's behaviour.

\begin{center}
\includegraphics[width=2.9in]{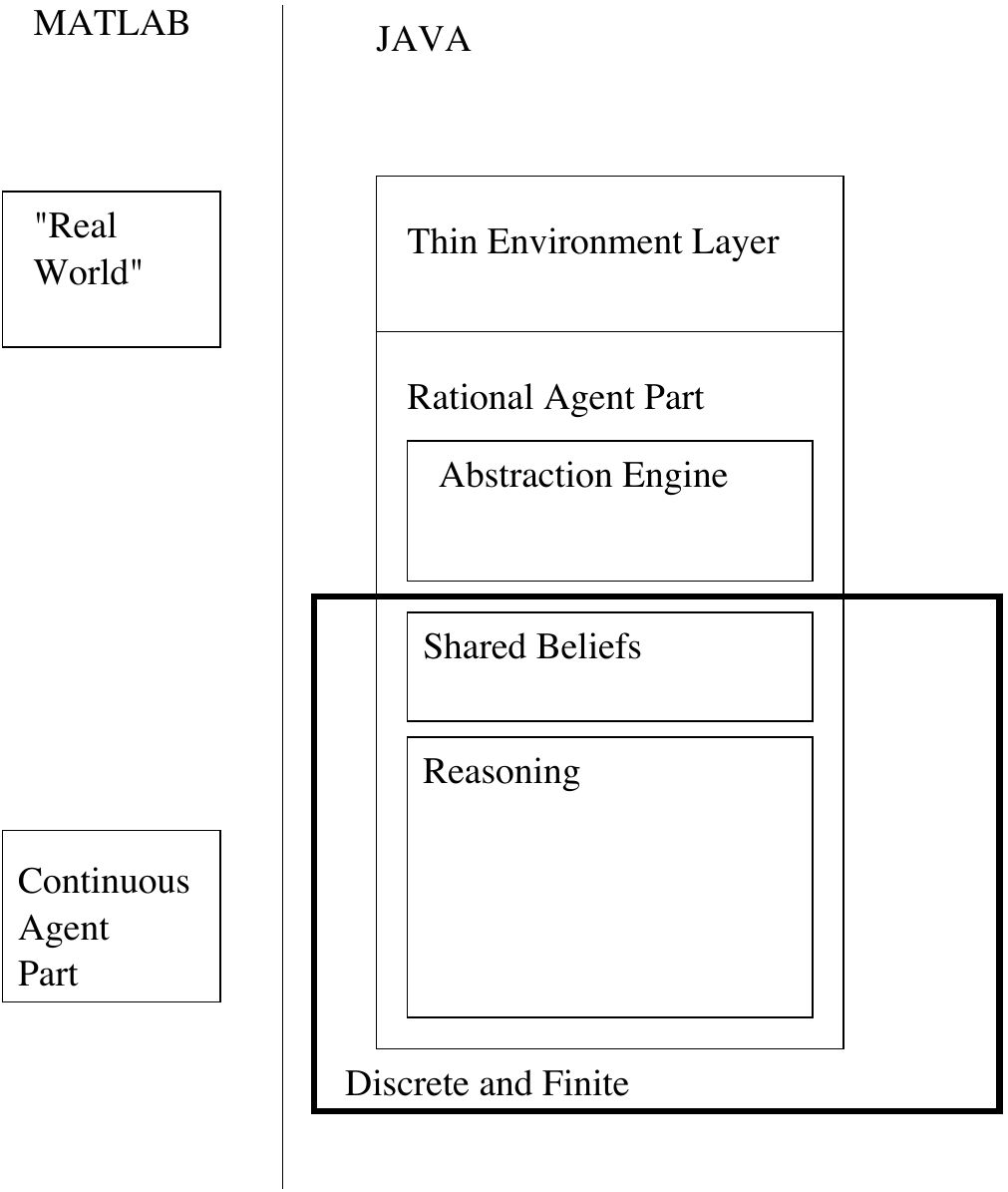}
\end{center}


\bibliographystyle{abbrv}
\bibliography{eass}

\end{document}